\documentclass[runningheads]{llncs}
\usepackage[T1]{fontenc}
\usepackage{graphicx}

\usepackage{amsmath,amssymb,amsfonts}
\usepackage{textcomp}
\usepackage{xcolor}

\usepackage{multirow}
\usepackage{tabularx}
\usepackage{subcaption}

\usepackage{algorithm}
\usepackage{algpseudocode}

\begin{document}
\title{FedMADE: Robust Federated Learning for Intrusion Detection in IoT Networks Using a Dynamic Aggregation Method}
\titlerunning{Robust Federated Learning for Intrusion Detection in IoT Networks}

\author{Shihua Sun \and Pragya Sharma \and
Kenechukwu Nwodo \and
Angelos Stavrou \and
Haining Wang
}
\authorrunning{S. Sun et al.}

\institute{
Virginia Tech, Arlington VA, USA \\
\email{\{shihuas,pragyasharma,nwodok,angelos,hnw\}@vt.edu} 
}

\maketitle              
\begin{abstract}
The rapid proliferation of Internet of Things (IoT) devices across multiple sectors has escalated serious network security concerns. This has prompted ongoing research in Machine Learning (ML)-based Intrusion Detection Systems (IDSs) for cyber-attack classification.
Traditional ML models require data transmission from IoT devices to a centralized server for traffic analysis, raising severe privacy concerns. To address this issue, researchers have studied Federated Learning (FL)-based IDSs that train models across IoT devices while keeping their data localized.
However, the heterogeneity of data, stemming from distinct vulnerabilities of devices and complexity of attack vectors, poses a significant challenge to the effectiveness of FL models. 
While current research focuses on adapting various ML models within the FL framework, they fail to effectively address the issue of attack class imbalance among devices, which significantly degrades the classification accuracy of minority attacks. 
To overcome this challenge, we introduce FedMADE, a novel dynamic aggregation method, which
clusters devices by their traffic patterns and aggregates local models based on their contributions towards overall performance.
We evaluate FedMADE against other FL algorithms designed for non-IID data and observe up to 71.07\% improvement in minority attack classification accuracy.
We further show that FedMADE is robust to poisoning attacks and incurs only a 4.7\% (5.03 seconds) latency overhead in each communication round compared to FedAvg, without increasing the computational load of IoT devices.

\end{abstract}
\section{Introduction}


The last decade has witnessed significant advancements in the Internet of Things (IoT) across various domains, including smart city infrastructures, e-health systems, and transportation networks~\cite{iot}. The widespread deployment of IoT devices expands the attack surfaces, and the inadequate security mechanisms of many IoT devices increase the vulnerability of these systems to cyber-threats. In response to these security issues, Intrusion Detection Systems (IDSs) have emerged as a critical defensive measure to protect IoT networks.
Typical IDSs primarily utilize two attack detection approaches: anomaly-based and signature-based~\cite{IDS_survey,signature}. The former method employs learning-based techniques to differentiate between benign and malicious traffic patterns, while the latter method inspects traffic against known attack signatures.

Prevalent Machine Learning (ML) based IDSs~\cite{ML3,few_shot,N-BaIoT,ML2,ML1} collect data from devices across the network and process it on a central server, thus compromising data privacy. Therefore, in recent years, Federated Learning (FL)~\cite{fedavg} has emerged as a distributed IDS solution that protects data privacy by decentralizing the training of ML models. 
These models are locally trained and updated at the client-end and are sent to a central server for the construction of a global model. 
However, FL-based IDSs face challenges due to data heterogeneity, caused by diverse device types, device vulnerabilities, complex attack vectors, and varying data collection environments~\cite{diot_2019}. The disparity in attack distribution across devices, which we refer to as device-specific class imbalance, results in poor classification accuracy, particularly for minority attack classes. In this paper, we define minority attacks as those that impact only a small subset of devices.
 For the scope of our work, we analyze the CICIoT2023 dataset~\cite{CICIoT2023} which is currently the most extensive dataset for IoT attack analysis. In this dataset, prevalent attacks such as Denial of Service (DoS), reconnaissance, and Mirai botnet attacks are observed across the majority of devices, whereas web-based and brute force attacks remain restricted to a limited number of devices. 
While strategies like undersampling and oversampling can alleviate the class data quantity imbalance in centralized environments, these approaches fall short in addressing the device-specific class imbalance prevalent in modern distributed IDS frameworks~\cite{smote}. 

Existing FL-based IDS solutions~\cite{fed-dbn_2023,iot-sn_2020,fl-lstm-gru_2022,fed-pca_2023,fl-botnet_2022,feco_2022} are inadequate in addressing the data heterogeneity problem of IoT networks. Among these works, \cite{iot-sn_2020,fed-pca_2023,fl-botnet_2022,feco_2022} split centralized datasets in a statistical manner with artificially constructed non-independent and identically distributed (non-IID) distributions (see Table~\ref{tab:FL-IDS}), which fail to reflect the real IoT network traffic distribution. Meanwhile, \cite{fed-dbn_2023,fl-lstm-gru_2022,fed-pca_2023} emphasize on integrating local models with different architectures to enhance their performance, but they neglect the impact of model divergence among clients in global model aggregation.

This paper presents FedMADE, a dynamic aggregation method designed for data heterogeneity management and \underline{m}inority \underline{a}ttack \underline{d}etection \underline{e}nhancement in FL-based IDSs. 
This method is implemented solely on the central server, without disturbing the local model update process, and therefore does not require additional computational resources from IoT devices.

FedMADE comprises two steps: device clustering and dynamic aggregation weight design~\footnote{The term ``aggregation weight'' refers to the coefficient associated with the local model during the aggregation process, distinct from the model weights. }.
In the clustering phase, local models that exhibit similar classification performance are grouped together as detailed here. 
This classification performance is represented using a class probability matrix (CPM), with each row representing the average probability vector (i.e. the output vector of the last layer after applying softmax function) of a class across all samples.
In practice, the server computes CPMs for each local model using a randomly-sampled minor auxiliary dataset for validation. Subsequently, the DBSCAN algorithm~\cite{dbscan} is applied to these matrices to facilitate model clustering. Each cluster is then treated as a single entity to calculate aggregation weights, which reduces the computational complexity involved in analyzing large-scale IoT networks.
For each cluster, we average the class probabilities from each model to generate cluster class probability matrices (CCPMs). 
We formulate the design of aggregation weights as an optimization problem aimed at minimizing the distance between the weighted sum of CCPMs and the identity matrix. The identity matrix represents the ideal CPM, in which each sample is classified into the correct class with 100\% confidence.
As a result, local models that exhibit consistently superior performance will be allocated larger coefficients, thereby increasing their influence on the global model.

Apart from the performance degradation issues in IoT networks, FL frameworks are also vulnerable to various adversarial attacks~\cite{gradient_inver,malware_det_2022,FL_private,scale-mia,targeted}. Prominent among these are poisoning attacks~\cite{malware_det_2022,targeted}, where a set of local IoT clients may be compromised to erroneously steer the FL model towards incorrect classification. In this adversarial environment, FedMADE can also serve as a robust aggregation method to protect FL-based IDSs from such attacks. By using CPMs as performance indicators, the server can identify suspicious models that adversely affect the optimization of aggregation weights.
Consequently, compromised models can be allocated minimal or zero coefficients to limit their effect on the global model. 

Our contributions are summarized below: 
\begin{itemize}
    \item We introduce a dynamic aggregation approach called FedMADE, that adaptively adjusts aggregation weights for FL-based IDS solutions. This approach selectively allocates larger weights to local models that demonstrate a consistent ability to accurately classify samples across all classes.
    \item We conduct a comprehensive evaluation of FedMADE, utilizing the CICIoT2023 dataset, for binary attack detection and multiclass classification scenarios and demonstrate its effectiveness in addressing device-specific class imbalance.
    FedMADE increases the classification accuracy of minority attacks by up to 71.07\%, with only a marginal 4.7\% (5.03 seconds) increase in round training time compared to FedAvg.
    \item Lastly, we examine the robustness of FedMADE against poisoning attacks in adversarial settings. 
\end{itemize}

\section{Background \& Related works}
In this section, we outline the background of FL and current FL-based IDSs for IoT networks, discuss the challenges of applying FL in IoT environments, and present existing FL methods designed to address data heterogeneity.

\begin{table*}[]
\centering
\caption{Overview of existing FL-based IDS methods, including their utilized datasets and data partition strategies.}
\label{tab:FL-IDS}
\resizebox{1\columnwidth}{!}{%
\begin{tabular}{l|l|l|l}
\hline
\textbf{Dataset: Collection Environment}      & \textbf{Method}    & \textbf{Partition}     & \textbf{Attacks}         \\ \hline\hline
Modbus: N/A    & FL-LSTM-GRU (2022)~\cite{fl-lstm-gru_2022}         & N/A                                                                       & DDoS, MITM     \\ \hline

\begin{tabular}[c]{@{}l@{}}NSL-KDD (2009)~\cite{NSL-KDD}:\\ Simulated U.S. Air Force LAN\end{tabular}                            & \begin{tabular}[c]{@{}l@{}}FeCo (2022)~\cite{feco_2022} \\ Fed-PCA (2023)~\cite{fed-pca_2023} \\ IoT-SN (2020)~\cite{iot-sn_2020}\end{tabular} & Statistics                                                                & DoS, Probe, R2L, U2R     \\ \hline
\begin{tabular}[c]{@{}l@{}}UNSW-NB15 (2015)~\cite{UNSW-NB15}:\\  Virtual network \end{tabular}                                      & IoT-SN (2020)~\cite{iot-sn_2020}                                                                                      & Statistics           & \begin{tabular}[c]{@{}l@{}}Fuzzers, Analysis, Backdoors, DoS, \\Exploits, Generic, Recon, Shellcode \\and Worms\end{tabular}     \\ \hline

\multirow{2}{*}{\begin{tabular}[c]{@{}l@{}}N-BaIoT (2018)~\cite{N-BaIoT}:\\ Realistic network with 9 IoT devices\end{tabular}}          
& FL-IoT-Malware (2022)~\cite{malware_det_2022}     & Device  & \multirow{2}{*}{BASHLITE, Mirai}                               \\ \cline{3-3}
& IoT-SN(2020)~\cite{iot-sn_2020}  & Statistics     &                                                                     \\ \hline
\begin{tabular}[c]{@{}l@{}}Bot-IoT (2018)~\cite{BoT-IoT}:\\ Virtual network  \end{tabular}                                 & FL-Botnet (2022)~\cite{fl-botnet_2022}                                                                                              & Statistics                                                                & DoS, DDoS, Recon, and Data theft      \\ \hline
\begin{tabular}[c]{@{}l@{}}{\"D}IoT dataset (2019)~\cite{diot_2019}:\\ Realistic network with 33 IoT devices\end{tabular}                       & {\"D}IoT (2019)~\cite{diot_2019}                                                                                           & Device                                                                    & \begin{tabular}[c]{@{}l@{}}Four attack states of Mirai malware: \\ pre-infection, infection, scanning, DoS\end{tabular}        \\ \hline

\begin{tabular}[c]{@{}l@{}}TON-IoT (2020)~\cite{TON}:\\ Network with 7 IoT and IIoT sensors
\end{tabular}                   & Fed-DBN (2023)~\cite{fed-dbn_2023}                                                                                        & IP addresses     & \begin{tabular}[c]{@{}l@{}}Scanning, DDoS, DoS, Ransomware, \\Backdoor, Injection,  XSS, \\Password Cracking, MITM\end{tabular} \\ \hline

\begin{tabular}[c]{@{}l@{}}\textbf{CICIoT2023 (2023):} \\ \textbf{Realistic IoT network with 105} \\\textbf{devices}\end{tabular}          & \textbf{FedMADE}        & \begin{tabular}[c]{@{}l@{}}Victim MAC \\ addresses by \\attack\end{tabular} & \begin{tabular}[c]{@{}l@{}}DoS, DDoS, Mirai, Recon, \\ Spoofing, Web-based, \\ Brute Force\end{tabular}                           \\ \hline
\end{tabular}
}
\vspace{-0.1in}
\end{table*}

\subsection{Federated Learning} ~\label{sec:fl}
An FL system consists of a central server and a set of distributed local clients that collaboratively develop a global model without sharing their data with the server or each other.
 The FL model training process is as follows: (1) the central server initializes a global model and distributes it to the local clients; (2) the clients update their local models based on captured traffic and then send their updated models back to the server; (3) the server aggregates the local models and broadcasts the updated global model to the clients for the next training round~\footnote{
In this paper, the terms ``training rounds'' and ``communication rounds'' are used interchangeably.}. Steps (2) and (3) are repeated until the FL model converges. 
 In most works, FedAvg~\cite{fedavg} is employed as the default aggregation method in step (3), which averages the local models with weights proportional to their dataset sizes.

\subsection{Existing FL-based IDSs for IoT Networks}
Nguyen et al.~\cite{diot_2019} introduced D{\"I}oT, a pioneering FL strategy for intrusion detection within IoT networks. This method utilizes diverse device-type-specific models to effectively manage the inherent heterogeneity among IoT devices.
Wang et al.~\cite{feco_2022} introduced FeCo, a representation learning method designed to create a new feature space. This space allows for the precise classification of benign and attack traffic, as benign traffic clusters tightly together, making it clearly distinguishable from attack traffic. 
Popoola et al.~\cite{fl-botnet_2022} focused on zero-day botnet attack detection, demonstrating the superior performance of their FL strategy over both centralized and localized detection methods.
Nguyen et al.~\cite{fed-pca_2023} incorporated Principal Component Analysis (PCA) into the FL framework, employing Grassmann manifold optimization to expedite the training process and reduce computation time.
The integration of diverse ML models into FL frameworks for IoT networks has been explored in several studies: \cite{iot-sn_2020} employed Deep Belief Networks (DBNs) and autoencoders; \cite{AMCNN-LSTM_2021} utilized Attention Mechanism-based Convolutional Neural Network Long Short-Term Memory (AMCNN-LSTM) models for anomaly detection; \cite{malware_det_2022} incorporated various architectures of Multi-Layer Perceptrons (MLPs) and autoencoders; \cite{fl-lstm-gru_2022} applied LSTM networks and Gated Recurrent Units (GRUs); and \cite{fed-dbn_2023} integrated Deep Neural Networks (DNNs) and DBNs. In addition, Liu et al.~\cite{AMCNN-LSTM_2021} introduced a gradient compression mechanism aimed at reducing the communication overhead associated with transmitting models between the server and local devices.
Rey et al.~\cite{malware_det_2022} 
discussed the vulnerabilities of FL models to poisoning attacks in the context of IDSs. 
Belarbi et al.~\cite{fed-dbn_2023} employed a pre-trained global model to facilitate the convergence of the global model during training.


Table~\ref{tab:FL-IDS} summarizes existing FL-based IDS solutions, the datasets employed, and their data partition strategies. 
The data partition strategies include `Statistics,' which denotes either IID or artificially designed non-IID distributions; `Device,' indicating datasets pre-split by device; `IP addresses,' describing allocations based on destination IP addresses; and `Victim MAC addresses by attack,' where traffic from each attack is distributed to local clients based on the corresponding victim devices (further details are provided in Section~\ref{sec:system_model}).

\textit{Challenges of deploying FL to IoT networks.} (1) Lack of realistic network traffic in the FL context: IDS datasets are typically collected centrally within the network, necessitating a realistic data splitting strategy that effectively replicates the intricate traffic patterns of real-world IoT networks. However, most existing works~\cite{iot-sn_2020,fed-pca_2023,fl-botnet_2022,feco_2022} split these centralized IDS datasets into either IID or artificially designed non-IID distributions, which do not adequately represent the true network traffic scenarios. 
(2) Traffic heterogeneity: Various devices are susceptible to different types of attacks. While the diversity of attack types may not significantly impact binary classification for attack detection, it substantially affects the accuracy of classification to more detailed attack classes. This issue impedes the accurate assessment of network conditions. Additionally, much of the current research~\cite{iot-sn_2020,malware_det_2022} aims to incorporate diverse ML models into the FL framework using the standard FedAvg aggregation, which neglects unique attack class imbalances among devices. 
(3) Latency and resource constraints: The adoption of FL into IoT networks must take into account the limitations of computational resources and memory. 
(4) Scalability: Considering the large scale of IoT networks, FL-based IDS solutions must be capable of accommodating a large number of participating devices.
(5) Robustness: Given the expansion of attack surfaces with FL, poisoning attacks can severely degrade the performance of FL-based IDSs by altering or injecting data into compromised IoT devices~\cite{poison}. Thus, a robust FL framework is crucial for practical IoT networks.

\subsection{FL with Data Heterogeneity}
Previous works design their methods to accommodate heterogeneous data by focusing on the three techniques:
(1) adjusting local model updates~\cite{scaffold,MOON,fedprox}, (2) loss re-weighting and data augmentation~\cite{ratio_loss,fedmix}, and (3) intelligent local model selection and aggregation~\cite{cluster_sample_2021,node_selection,mindfl}.
FedProx~\cite{fedprox} incorporates a proximal term into the loss function of local clients, which limits the divergence between updated local models and the global model.
SCAFFOLD~\cite{scaffold} measures client drift through the difference between update directions of the server and the clients, subsequently adjusting for this drift during the local model update.
MOON~\cite{MOON} leveraged contrastive learning to minimize the discrepancy between representations derived from the global and local models. Wang et al.~\cite{ratio_loss} developed a new loss function, called Ratio Loss, which adjusts the contributions of different training samples to emphasize those from minority classes. 
Yoon et al.~\cite{fedmix} approximated the impact of Global Mixup, a linear interpolation-based data augmentation technique, by assuming that clients could exchange averaged data with others.
Fraboni et al.~\cite{cluster_sample_2021} developed a client sampling strategy aimed at reducing the variability of selected models during training and improving client representations.
Wu et al.~\cite{node_selection} increased the client sampling probability for those with higher similarity between their local gradients and the global gradient.

\begin{table*}[t]
\centering
\caption{CICIoT2023 Dataset Statistics: number of samples in each class, class size ratios, and victim device count per class.}
\label{tab:cic-iot}
\resizebox{\columnwidth}{!}{
\begin{tabular}{|c|c|c|c|c|c|c|c|c|}
\hline
         & Benign    & DDoS        & DoS        & Mirai     & Recon    & Spoofing & Web-based & Brute Force \\ \hline
\#data samples & 1,098,195 & 33,984,560 & 8,090,738 & 2,634,124 & 354,565 & 486,504 & 24,829   & 13,064   \\ \hline
Ratio (\%)      & 2.35         & 72.79          &  17.33        & 5.64        &0.76        & 1.04       & 0.05         & 0.03          \\ \hline
\#victims      & 63        & 63          & 54         & 61        & 63       & 58       & 3         & 5          \\ \hline
\end{tabular}
}
\vspace{-0.1in}
\end{table*}
\section{System model} \label{sec:system_model}

\subsection{Dataset and Data Partition} For this work, we utilize the latest IDS dataset, CICIoT2023~\cite{CICIoT2023}, which was collected from an extensive network topology of 105 real IoT devices, including smart home devices, cameras, sensors, and micro-controllers. This network comprises 67 devices involved in various attacks (63 victims and 4 attackers) and 38 Zigbee and Z-Wave devices. Moreover, the dataset covers the most comprehensive range of attacks targeting IoT networks to date, including 33 attacks divided into 7 broad categories: DDoS, DoS, Mirai, Recon, Spoofing, Web-based, and Brute Force. The packet capture (pcap) files are processed using the DPKT library~\cite{dpkt} to extract 47 features~\footnote{For more details on the extracted features, please refer to~\cite{CICIoT2023}.}, saved in CSV files with corresponding labels. 
In contrast, alternative IDS datasets~\cite{BoT-IoT,N-BaIoT,TON,UNSW-NB15,NSL-KDD} from IoT environments are often derived from small-scale virtual networks that lack realistic device configurations, or they cover only a limited set of attacks. These limitations render such datasets inadequate for a thorough evaluation of the generalization and scalability of IDS solutions. 
Table~\ref{tab:cic-iot} presents the data distribution of CICIoT2023. While the CSV files lack the source and destination IP addresses of traffic flows, this dataset provides a list of victims for each type of attack, including their MAC addresses. Consequently, in our FL framework, we designate each client as an IoT device identified by a MAC address and distribute network traffic from each class to the corresponding clients based on the victim list.

\subsection{FL-based IDS}
As explained in Section~\ref{sec:fl}, the FL system consists of two fundamental categories of components: multiple distributed clients and a central server.
In the deployment of an FL-based IDS in IoT networks, the local model of each device is managed by its respective gateway, which connects the device to the Internet, and the central server is responsible for aggregating these local models. 
For the CICIoT2023 scenario, we developed an FL system with 63 clients, where each client corresponds to a victim IoT device. 
Local models are updated using gradient descent on the client side and FedMADE is employed on the server side to dynamically aggregate these local models.
During the FL training process, clients can choose to participate in each communication round based on the sporadic availability of IoT devices influenced by factors such as CPU, memory, battery capacity, and network connectivity~\cite{fedprox}.

\section{Proposed Method: FedMADE}

\begin{figure*}[t]
\centering
\centerline{\includegraphics[trim={0.2cm 1.1cm 12cm 3.4cm},clip, width=0.7\linewidth]{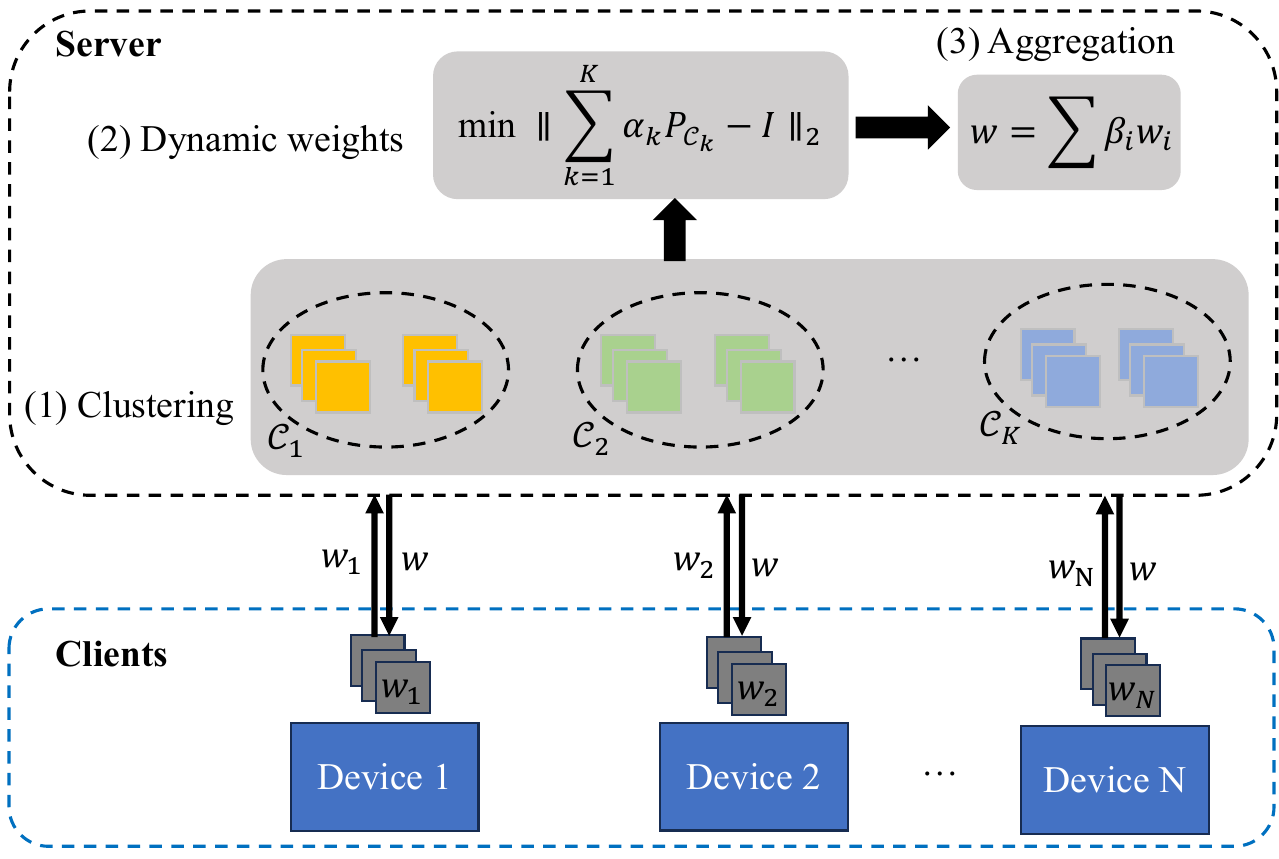}}
\caption{The visualization of FedMADE-based IDS.}
\label{fig:framework}
\end{figure*}


The primary cause of poor classification of minority attacks is the heterogeneous distribution of attack traffic. For instance, in the CICIoT2023 dataset, web-based attacks affect only 3 out of 67 devices. Consequently, with the use of FedAvg, knowledge about web-based attacks gets diminished during the aggregation process.
To address this issue, we propose FedMADE which incorporates a dynamic allocation of aggregation weights to different clients, prioritizing those with minority attack representations. This approach is designed to ensure robust classification performance across a diverse set of attack types. 

We formulate the design of aggregation weights as an optimization problem. In case of large-scale IoT networks, the scalability of our IDS solution can get limited with the increased number of devices. Thus, we propose a clustering of clients by the central server to reduce computation complexity. 
As illustrated in Figure~\ref{fig:framework}, the key steps in FedMADE are (1) Device Clustering: Upon receiving the local models, the server groups the clients into clusters based on their classification performance, (2) Weight Calculation: The server calculates aggregation weights for each cluster, with clients in the same cluster receiving identical weights, and (3) Model Aggregation: The local models are aggregated together with these dynamically assigned weights.
 The FedMADE approach is summarized in Algorithm~\ref{alg:FedMADE}. 
 
 In this section, we first provide an overview of the local model training process at the clients in Section~\ref{sec:local_training}. Further, we comprehensively detail the device clustering and dynamic aggregation approach in Sections~\ref{sec:clustering} and~\ref{sec:aggregation_weight}.

\subsection{Local Model Training} \label{sec:local_training}
During the $r$-th training round, clients receive the global model $w^{(r)}$ and update their local models following the formula ${w}^{(r)}_i = {w}^{(r)} -\eta \nabla L\left({w}^{(r)}, D_i\right)$, where $\eta$ is the learning rate, $D_i$ denotes the local dataset of the $i$-th client, and $L$ represents the loss function. These local model updates are conducted iteratively over several epochs, depending on the available resources of the IoT devices, such as battery capacity and CPU utilization.

\begin{algorithm}[t]
\caption{FedMADE}\label{alg:FedMADE}
\begin{algorithmic}[1]

\For {each round $r=1,...,R$}
    \State {Sample a set of clients $U$ according to the client sampling rate $\gamma$}
    \State {Broadcast $w^{(r)}$ to all clients in $U$, and do local updates} 
    \State {Collect local models \{$w^{(r)}_{1},...,w^{(r)}_{|U|}$\} from clients}

\For {each local model $w^{(r)}_{i}$}
\State {Compute class probability matrix $P_{w^{(r)}_{i}}$ according to Equation~\ref{eq:class_probability}}
\EndFor 
\State $\mathcal{C} \gets \operatorname{DBSCAN}( \{ P_{w^{(r)}_{1}},...,P_{w^{(r)}_{|U|}}\} ) $ 
\For {for each cluster $\mathcal{C}_k$}
\State $P_{\mathcal{C}_k} \gets \frac{1}{|\mathcal{C}_k|}\sum_{w_i^{(r)} \in \mathcal{C}_k}P_{w_i^{(r)}}$
\EndFor 
\State $\alpha_k \gets \text{argmin} \|\sum_{k=1}^{K} \alpha_{k} P_{\mathcal{C}_k}-I\|_2$
\State $\beta_{i} \gets \frac{\alpha_k}{\lvert \mathcal{C}_k \rvert}$ if $w_i^{(r)} \in \mathcal{C}_k$ for each $w_i^{(r)}$
\State $\beta_{i} \gets \frac{\beta_{i}}{\sum_{i}^{|U|} \beta_{i}}$ for each $\beta_{i}$
\State $w^{(r+1)} \gets \sum_{i}^{|U|} \beta_i w_i^{(r)} $
\EndFor 
\end{algorithmic}

\end{algorithm}

\subsection{Client Clustering} \label{sec:clustering}
Consider the local model in the $i$-th client from the $r$-th training round, denoted as $w_i^{(r)}$. The associated mapping function from the feature space to the output space is represented as $f_{w_i^{(r)}}$. 
Given an input sample $x$, the output probability vector is denoted as $p(x)=f_{w_i^{(r)}}(x)$. Each component of $p(x)$ represents the likelihood that the input is classified into a specific class, with the sum of all components equaling 1. To represent the capability of local models to differentiate between different attack classes, we introduce a metric called Class Probability Matrix (CPM). This metric is derived by averaging the output probabilities of all samples within each class. Specifically, for class $c$, the class probability vector is calculated as $\bar{P_c} = \frac{1}{n_c}\sum_{j=1}^{n_c}p(x_{c,j})$, where $n_c$ represents the number of data samples in class $c$, and $x_{c,j}$ is the $j$-th data sample of that class. Finally, by concatenating the class probability vectors for all classes, we construct the CPM for a local model $w_i^{(r)}$:
\begin{equation} 
     P_{w_i^{(r)}} = [\bar{P_1}, \bar{P_2}, \cdots, \bar{P_{N_C}}],
     \label{eq:class_probability}
\end{equation}
where $N_C$ is the number of classes.

In each training round, when the server collects local models from the local client gateways, it utilizes an auxiliary dataset to calculate the CPM for each local model. For our experiments, we construct a small auxiliary dataset of ten samples from each class.  
We denote the set of clients that participate in the $r$-th training round as $U$, and the collected local models from the participants as ${w^{(r)}_{1},...,w^{(r)}_{|U|}}$, where $|U|$ is the number of participants in this training round.
Next, the server applies the DBSCAN~\cite{dbscan} clustering technique to the obtained CPMs, and groups their corresponding clients into different clusters. 
DBSCAN, a density-based algorithm, identifies core points with nearby neighbors and does not require specifying the number of clusters. This technique is suitable for our case, as CPMs from clients with similar classification performance naturally form compact and dense clusters. 
Using this technique, the clients which have similar class-specific performance can be grouped in the same cluster. Ultimately, we obtain the set $\mathcal{C}$ of $K$ client-clusters, $\mathcal{C} = \{\mathcal{C}_1, \mathcal{C}_2, \cdots,  \mathcal{C}_K\}$.

\subsection{Dynamic Aggregation Weight Calculation} \label{sec:aggregation_weight}
We formulate the dynamic assignment of aggregation weights as an optimization problem based on the local CPMs. When applied to large-scale networks, the computation complexity of our method can grow significantly if each client participates individually in solving the optimization problem. Hence, we view each client-cluster as a single entity and assign the same aggregation weights to all the clients in the same cluster.
For each cluster, we average the CPMs obtained from all local models in that cluster to get the averaged Cluster Class Probability Matrix (CCPM). The averaged CCPM for the cluster $\mathcal{C}_k$ is represented as
\begin{equation}
    P_{\mathcal{C}_k} = \frac{1}{|\mathcal{C}_k|}\sum_{w_i^{(r)} \in \mathcal{C}_k}P_{w_i^{(r)}}.
\end{equation}
The averaged CCPMs reflect the classification performance of local models in each cluster. The ultimate objective of the global model is to achieve consistently strong performance across all attack classes, including minority ones. To this end, the problem is designed to align the aggregated CCPMs with the identity matrix $I$. Here, $I$ represents the ideal classification scenario, with each row representing a distinct class and the probability of correctly classifying an input into its corresponding class being 100\%. In practice, the optimization problem aims to constrain the $L_2$ distance between the aggregated CCPMs and the identity matrix within a tolerance limit of $\epsilon$. To achieve this, the optimization expression can be formulated as
\begin{equation} \label{eq:optimization}
\min \|\sum_{k=1}^{K} \alpha_{k} P_{\mathcal{C}_k}-I\|_2,
\end{equation}
with the constraint that $\alpha_{k}$, which represents the aggregation weight of client cluster $\mathcal{C}_k$, is non-negative. Then, for each client, the client weight is the cluster weight divided by the number of clients in that cluster. For a local model $w_i^{(r)}$ of the $i$-th client, belonging to a cluster  $\mathcal{C}_k$, the aggregation weight is 
\begin{equation}
    \beta_{i} = \frac{\alpha_k}{\lvert \mathcal{C}_k \rvert}.
\end{equation}
Next, the aggregation weights are normalized so that their sum equals 1.
Intuitively, a local model that exhibits higher probabilities for accurate class predictions and maintains consistent performance across all classes receives greater coefficients. 
Finally, with the dynamic weights, the new aggregated global model for the next training round can be expressed as
\begin{equation}
    w^{(r+1)} = \sum_{i}^{|U|} \beta_i w_i^{(r)}.
\end{equation}

\section{Evaluation}
In this section, we first outline our experimental setup, followed by a comparison of FedMADE with other FL methods in terms of attack detection and classification. Finally, we analyze the robustness and latency of our method to examine the practicality and feasibility of implementing FedMADE.

\subsection{Experimental Setup} 

\textbf{Data Preprocessing.} The CICIoT2023 dataset is divided into training and testing subsets, containing 80\% and 20\% of the samples, respectively. The input features are normalized using min-max scaling. 
To alleviate the class imbalance issue in local traffic, the Synthetic Minority Oversampling Technique (SMOTE)~\cite{smote} is employed to oversample classes, excluding DoS, DDoS and Mirai attacks. Specifically, the number of samples in web-based and brute force attacks is quadrupled, while the number of samples from other traffic classes is doubled. This configuration is designed to improve the representation of minority classes without causing the model to become overfitted to them. SMOTE generates synthetic samples for minority classes by interpolating between several close examples in the feature space.

\textbf{FL System.} We employ two base models in the FL system: a Convolutional Neural Network (CNN) and a Fully Connected Neural Network (FCNN). The CNN model is structured with two convolutional layers followed by one dense layer.  
The FCNN model consists of three dense layers, each followed by a dropout layer. The more detailed architectures are provided in Appendix~\ref{appendix:arch}.
In each communication round, a subset of local clients is randomly selected to participate, with client sampling rate $\gamma \in \{0.5,1\}$. 
While $\gamma = 1$ expresses the ideal scenario where all local models participate in each communication round,  
$\gamma = 0.5$ demonstrates a more practical scenario where devices with minority attacks do not consistently participate in each training round\footnote{The $\gamma$ value can be adjusted based on real IoT network configurations.}. 
In each communication round, participating clients update their local models for two epochs using the Adam optimizer set at a learning rate of $5\times10^{-4}$. After local updates, these models are sent back to the central server. The server then generates a new global model and evaluates it using a validation dataset. If the new model achieves higher classification accuracy than the previous ones, it is retained as the best model until it is surpassed by another model with superior accuracy. Eventually, the best model is deployed in all IoT devices for attack detection or classification.
To calculate aggregation weights, we use the Adam optimizer with a learning rate of 0.01 to solve the optimization problem.

\textbf{Evaluation Metrics.} To evaluate the overall detection performance, we employ metrics such as precision, recall, F1 score, and accuracy. Additionally, to assess the model’s performance on each minority class, we employ per-class accuracy to overcome the effect of class size in the testing dataset. Precision is defined as $\frac{TP}{TP+FP}$, where TP and FP denote the number of true positives and false positives, respectively. Recall is calculated as $\frac{TP}{TP+FN}$, with FN indicating the number of false negatives. The F1 Score is computed as $2\times \frac{Precision \times Recall}{Precision + Recall}$. Overall accuracy is determined by the ratio of correctly classified samples to the total number of samples, while per-class accuracy quantifies the proportion of correctly classified samples within each specific class.

\subsection{Performance of FedMADE}

We begin by evaluating FedMADE's ability to detect attacks through binary classification. Following that, we compare FedMADE with FedAvg and two other well-known methods that are designed for non-IID data—FedProx~\cite{fedprox} and SCAFFOLD~\cite{scaffold}—on attack classification into multiple classes.

\textbf{Attack Detection (Binary classification).} 
The main focus of our approach is on improving multi-class classification instead of binary classification, due to the fact that device-specific class imbalance becomes less significant when all attack types are grouped into a single class. However, we still present binary classification results for FedMADE and FedAvg to show their effectiveness in attack detection. Tables~\ref{tab:binary} and~\ref{tab:binary_gamma05} show the precision, recall, F1-score, and accuracy of using CNN and FCNN models. We observe that the performance of FedMADE is very similar to that of FedAvg, with both methods demonstrating strong attack detection capabilities and achieving F1 scores greater than 0.99, despite a marginal reduction in recall (up to 0.0023) in some cases.
This is largely because minority attack classes, such as web-based and brute force attacks, make up only 0.08\% of all traffic flows, whereas major attacks like DoS/DDoS and Mirai constitute 95.76\% of the traffic. Given the small proportion of minority attack classes, their impact on the overall performance of binary attack detection is negligible. Furthermore, rather than assuming all clients are honest, we illustrate the robustness of FedMADE to poisoned clients in Section~\ref{sec:robustness}.

\begin{table}[t]
\centering
\caption{Performance on attack detection with $\gamma=1.0$.}
\label{tab:binary}
\begin{tabular}{|c|c|c|c|c|c|}

\hline
Base Model           & Aggregation & Precision & Recall & F1 Score     & Accuracy \\ \hline
\multirow{2}{*}{CNN} & FedAvg      & 0.9982    & 0.9947 & 0.9965 & 0.9931   \\ \cline{2-6} 
                     & FedMADE     & 0.9990    & 0.9931 & 0.9960 & 0.9923   \\ \hline

\multirow{2}{*}{FCNN} & FedAvg      & 0.9997    & 0.9913 & 0.9955 & 0.9912   \\ \cline{2-6} 
                     & FedMADE     & 0.9998    & 0.9908 & 0.9953 & 0.9908   \\ \hline

\end{tabular}
\vspace{-0.1in}
\end{table}

\begin{table}[t]
\centering
\caption{Performance on attack detection with $\gamma=0.5$.}
\label{tab:binary_gamma05}
\begin{tabular}{|c|c|c|c|c|c|}

\hline
Base Model           & Aggregation & Precision & Recall & F1 Score    & Accuracy \\ \hline
\multirow{2}{*}{CNN} & FedAvg      & 0.9983    & 0.9948  & 0.9965  & 0.9932   \\ \cline{2-6} 
                     & FedMADE     & 0.9985    & 0.9945  & 0.9965  & 0.9932   \\ \hline

\multirow{2}{*}{FCNN} & FedAvg     & 0.9989   & 0.9940 & 0.9965 & 0.9931   \\ \cline{2-6} 
                     & FedMADE     & 0.9997   & 0.9917 & 0.9957 & 0.9916   \\ \hline

\end{tabular}
\end{table}

\begin{figure*}[t]
    \centering
    \begin{subfigure}[b]{1\textwidth}
        \centering
        \includegraphics[width=\textwidth]{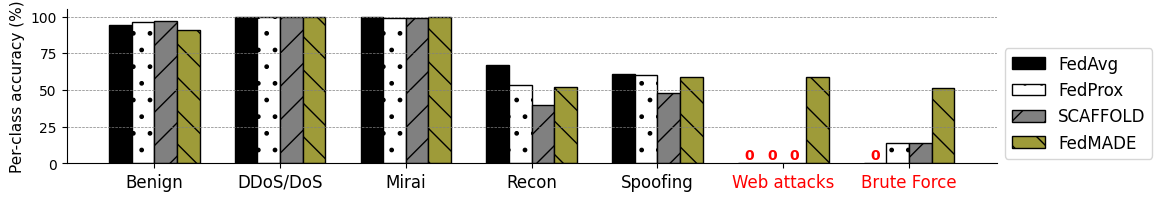}
        \caption{CNN model}
    \end{subfigure}
    \hfill 
    \begin{subfigure}[b]{1\textwidth}
        \centering
        \includegraphics[width=\textwidth]{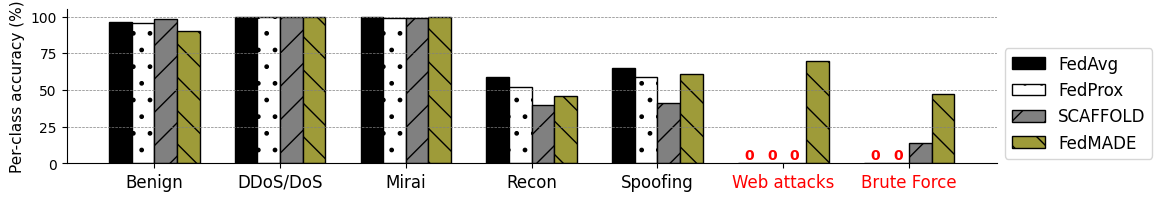}
        \caption{FCNN model}
    \end{subfigure}
    \caption{Comparison of performance across different FL methods for each class with a client sampling rate of 1.0. `0' denotes the FL model fails to correctly classify any sample from that class.}
    \label{fig:minority}
\end{figure*}

\begin{figure}[t]
    \centering
    \begin{subfigure}[b]{0.496\textwidth}
        \centering
        \includegraphics[width=\textwidth]{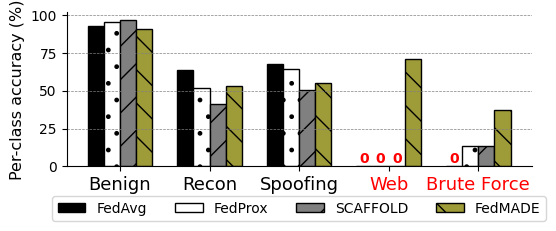}
        \caption{CNN model}
    \end{subfigure}
    \begin{subfigure}[b]{0.496\textwidth}
        \centering
        \includegraphics[width=\textwidth]{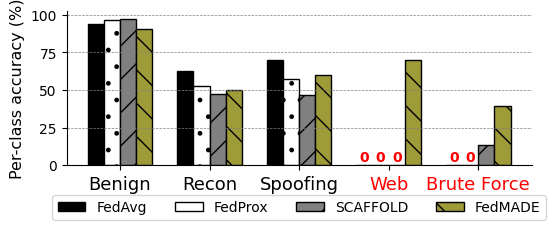}
        \caption{FCNN model}
    \end{subfigure}
    \caption{
Performance comparison at a client sampling rate of 0.5, excluding DDoS/DoS and Mirai attacks, as each FL method consistently achieves over 99\% accuracy for these attacks.}
    \label{fig:minority_gamma05}
    \vspace{-0.1in}
\end{figure}

\textbf{Attack Classification (Multi-class classification).} 
Considering the significant class size differences between different attack types, we use per-class accuracy to demonstrate the model's classification performance on each type of traffic. Figures~\ref{fig:minority} and~\ref{fig:minority_gamma05} compare the proposed FedMADE method with FedAvg, FedProx, and SCAFFOLD. When $\gamma$ is set to 1.0, FedMADE improves the classification accuracy for minority attack classes, including web-based and brute force attacks. Specifically, the classification accuracy of web-based attacks increases to 58.5\% and 69.5\%, using the CNN and FCNN models, respectively. For brute force attacks, the accuracy increases to 51.4\% with CNN and 47.1\% with FCNN. 
The improvement can be attributed to FedMADE's ability to direct the global model's attention towards local models that consistently demonstrate strong performance across all classes. In contrast, other FL methods either fail to classify minority attacks or achieve an accuracy of only up to 13.8\%. 
FedMADE exhibits a slight decrease in accuracy for classifying benign traffic and reconnaissance attacks, which is attributed to the inherent limitations of the base models (i.e., CNN and FCNN) in distinguishing between benign traffic, reconnaissance, web-based, and brute force attacks, as demonstrated in the original paper on the CICIoT2023 dataset~\cite{CICIoT2023}. These decreases are within an acceptable range, given the significant improvement in classifying minority attacks, leading to a more accurate diagnosis of network conditions.
Similarly, in a realistic setting with $\gamma$ set to 0.5, FedMADE increases the accuracy of minority attack classification to 71.07\%. In contrast, while FedProx and SCAFFOLD maintain accuracy for benign traffic, they severely impact the classification of reconnaissance and spoofing attacks as compared to FedMADE and do not improve the accuracy of minority attack classification.

\subsection{Robustness} ~\label{sec:robustness}

\begin{figure}[t]
    \centering
    \begin{subfigure}[b]{0.8\textwidth}
        \centering
        \includegraphics[width=\textwidth]{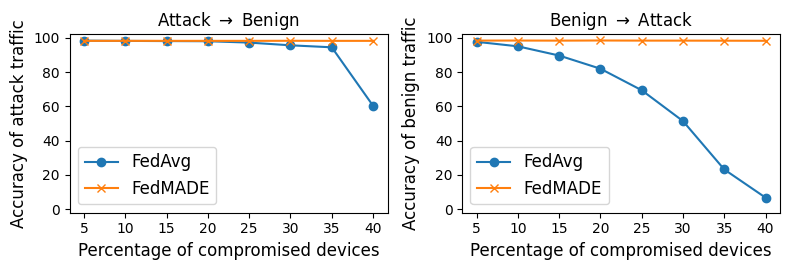}
        \caption{Data poisoning attacks}
    \end{subfigure}
    \hfill 
    \begin{subfigure}[b]{0.8\textwidth}
        \centering
        \includegraphics[width=\textwidth]{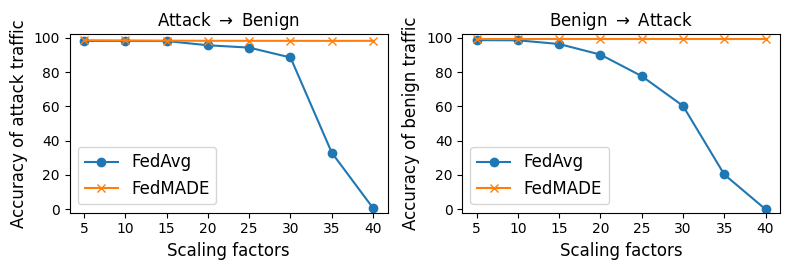}
        \caption{Model poisoning attacks}
    \end{subfigure}
    \caption{Per-class accuracy under label flipping and model poisoning attacks. `Attack $\rightarrow$ Benign' and `Benign $\rightarrow$ Attack' represent scenarios where attack traffic is labeled as benign, and benign traffic is labeled as attack, respectively.}
    \label{fig:robustness}
\end{figure}

Instead of assuming all clients are honest, we consider some clients are compromised in adversarial settings.
The attacker's goal is to mislead the FL model into incorrectly classifying incoming traffic. Following the attack scenarios targeting attack detection models outlined in~\cite{malware_det_2022,targeted}, we study two categories of poisoning attacks: data poisoning, where attackers modify the labels of the traffic data, and model poisoning, where the training process of the local models is manipulated. For both attacks, we explore two scenarios: attack label flipping, where attack traffic labels are altered to benign, and benign label flipping, where benign traffic labels are modified to attack.
The details of the poisoning attacks are as follows. (1) For data poisoning attacks, the percentage of compromised devices varies from 5\% to 40\%. While fewer compromised devices are more practical, we extend our analysis to include up to 40\% compromised clients to assess the resilience of our aggregation method under severe adversarial conditions. (2) For model poisoning attacks, we limit the percentage of compromised clients to 5\% and scale up the malicious models using different scaling factors.  Specifically, compromised clients upload model weights scaled by a factor, $\lambda w$, rather than the original weights $w$, to the server, where $\lambda$ ranges from 5 to 40 in evaluation.

Figure~\ref{fig:robustness} presents a performance comparison of FL-based IDSs using the CNN model under data and model poisoning attacks. For data poisoning, as the percentage of compromised devices increases, the per-class accuracy of victim classes declines with the FedAvg method. 
 In contrast, IDS performance using FedMADE remains stable and is not significantly affected by an increase in the number of compromised devices. For instance, in a scenario with 35\% compromised devices, the accuracy of attack and benign traffic classified using FedAvg decreases to $94.44\%$ and $23.22\%$ (Figure~\ref{fig:robustness}(a)), respectively. 
 However, with FedMADE, these values are observed to be $98.20\%$ and $98.35\%$ respectively. This robustness stems from the optimization mechanism where poisoned local models receive negligible aggregation weights due to the deviation of their CPMs from those of benign clients, consequently reducing their impact on the global model. We observe similar behaviour in case of model poisoning attacks (Figure~\ref{fig:robustness}(b)), bolstering our claim that FedMADE is more robust as an aggregation method compared to FedAvg within the context of IDS deployments in IoT networks.

\subsection{Latency Analysis}
\begin{table}[t]
\centering
\caption{Latency comparison. `Ratio' refers to the relative increase in each method’s round training time compared to that of FedAvg.}
\label{tab:latency}
\begin{tabular}{|c|cc|cc|}
\hline
\multirow{2}{*}{} & \multicolumn{2}{c|}{$\gamma$ = 0.5}                                                    & \multicolumn{2}{c|}{$\gamma$ = 1.0}                                                                    \\ \cline{2-5} 
                  & \multicolumn{1}{c|}{Round Training Times (s)} & Ratio & \multicolumn{1}{c|}{Round Training Times (s)} & Ratio \\ \hline
FedAvg            & \multicolumn{1}{c|}{107.46}    & -       &             \multicolumn{1}{c|}{221.82}     & -       \\ \hline
FedProx           & \multicolumn{1}{c|}{132.51}   & 23.3\%       & \multicolumn{1}{c|}{269.60}    &21.5\%   \\ \hline
SCAFFOLD          & \multicolumn{1}{c|}{161.38}  & 50.2\%       & \multicolumn{1}{c|}{333.78}    & 50.5\%       \\ \hline
FedMADE           & \multicolumn{1}{c|}{112.49}   & 4.7\%       & \multicolumn{1}{c|}{227.66}        & 2.6\%       \\ \hline
\end{tabular}
\end{table}
For the implementation of any FL-based IDS methods in practical, large-scale IoT network environments, it is essential to conduct an analysis of their latency in the distributed training process.
The factors contributing towards latency of FedMADE are (1) calculation of CPMs for local models (Equation~\ref{eq:class_probability}), and (2) minimization of the distance between the weighted sum of CPMs and the identity matrix (Equation~\ref{eq:optimization}).
It is worth noting that all operations associated with FedMADE are conducted on the central server, preventing any additional computational burden on the IoT devices.
Given the negligible differences observed in convergence rates with respect to training rounds (refer to Appendix~\ref{appendix:training}), we utilize round training times to analyze the latency overhead incurred by each FL method, as presented in Table~\ref{tab:latency}.
Each communication round includes the server broadcasting the global model to clients, clients training and uploading their local models, and the server aggregating these models to generate the new global model. With client sampling rates of $\gamma$ at 0.5 and 1.0, the process involves 31 and 63 local models in each round, respectively. Our experiments are conducted on a server equipped with an AMD EPYC 7763 Processor with an 8-core CPU, an NVIDIA A100 GPU, and Ubuntu 22.04.3 LTS. We sample 10\% of the training data from the CICIoT2023 dataset for performing this experiment.

As observed in Table~\ref{tab:latency}, the round training times of FedMADE show slight increases of 4.7\% and 2.6\% compared to FedAvg, for client sampling rates of 0.5 and 1, respectively. Given the gains in performance and robustness, these latency overheads are within an acceptable range. In contrast, the training times for FedProx and SCAFFOLD exhibit significantly large increases of 21.5\% and 50.2\%. Additionally, both methods involve modifications to the local training process and demand additional computational resources from IoT devices. Therefore, they are less suited for IoT networks, where devices are resource-constrained.

\subsection{Privacy Analysis}
To compute local CPMs, FedMADE requires only a small validation dataset consisting of ten samples per class, which is negligible in comparison to the extensive training dataset that comprises millions of samples. Moreover, it is a standard practice to validate any ML model before deploying it in practical settings. Thus, our method barely affects the privacy of devices in IoT networks. This is in contrast to other FL methods that compromise privacy by directly sharing local data distributions (e.g., class sizes) with the server~\cite{Fedsld}, distributing a proportion of the total data among other clients to mitigate local model weight divergence~\cite{FL_noniid}, or transmitting local feature statistics to the central server and other clients~\cite{FedPAC}.

\section{Discussion}
In our evaluations, we observe a slight decrease in accuracy for some classes, along with the improvement in minority attack classification. This is due to the limitations of local models in effectively differentiating between these minority attacks and some other traffic types, such as benign and reconnaissance traffic. To address this, we tested various feature selection methods including feature variance, information gain, recursive feature elimination, and tree-based approaches, as well as feature extraction techniques such as Principal Component Analysis (PCA) and Autoencoder~\cite{feature_selection}. However, these popular methods do not substantially enhance the classification among benign, web-based, brute force, and reconnaissance attacks. Advanced feature engineering approaches are essential to improve classification performance of local models; however, they are beyond the scope of this paper.
Furthermore, FedMADE can be combined with various advanced ML models for IDSs in IoT networks.

\section{Conclusion}
This paper adapts the FL framework to function as a distributed IDS in IoT networks for preserving privacy of IoT clients.
We introduced FedMADE, a dynamic aggregation method designed to handle traffic heterogeneity across IoT devices and enhance the detection capabilities for minority attacks. 
FedMADE shows strong performance in both detecting and classifying attacks, as demonstrated by its results with the CICIoT2023 dataset.
Moreover, our findings indicate that FedMADE is a robust and secure aggregation method, showing resilience against poisoning attacks in adversarial settings. 
FedMADE incurs a negligible latency overhead during training without imposing additional computational demands on local IoT devices. 
In conclusion, FedMADE offers a practical IDS solution for IoT networks, characterized by its robustness, scalability, low latency, and superior classification performance. 

%

\appendix
\section{Model Architecture} \label{appendix:arch}
The detailed architectures of the base models are provided in Tables~\ref{tab:arch_cnn} and ~\ref{tab:arch_fcnn}.

\begin{table}[h]
\centering
\vspace{-0.3in}
\begin{minipage}{.5\linewidth}
\caption{CNN Model architecture.}
\label{tab:arch_cnn}
\centering
\begin{tabular}{|c|c|c|}
\hline
Type               & Kernel & Shape \\ \hline
Convolution + Relu &    (3,3)    &  (64,47)     \\ \hline
Convolution + Relu  &   (3,3)     &    (128,47)   \\ \hline
Flatten   &    -    &    (6016)  \\ \hline
Dense + Softmax &  - &  (7) \\ \hline
\end{tabular}
\end{minipage}%
\begin{minipage}{.5\linewidth}
\centering
\caption{FCNN Model architecture.}
\label{tab:arch_fcnn}
\begin{tabular}{|c|c|}
\hline
Type             & Shape \\ \hline
Dense + Relu      &  (50)     \\ \hline
Dense + Relu        &    (25)   \\ \hline
Dense + Softmax    &  (7) \\ \hline
\end{tabular}
\end{minipage}
\vspace{-0.1in}
\end{table}

\section{Supplemental Results} 
\label{appendix:training}
\begin{figure*}[h]
\centerline{\includegraphics[width=1.0\linewidth]{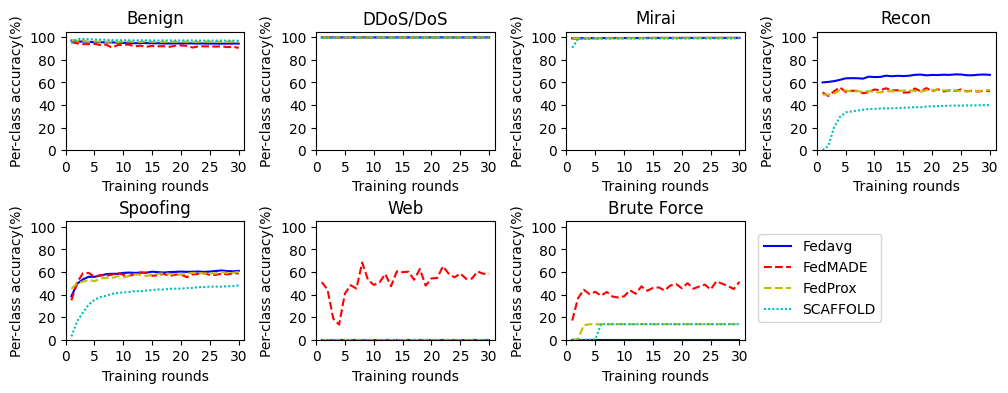}}
\caption{Curves of per-class accuracy across training rounds with $\gamma$ being 1.0.}
\label{fig:training_cnn}
\end{figure*}

\begin{figure*}[h]
\centerline{\includegraphics[width=1.0\linewidth]{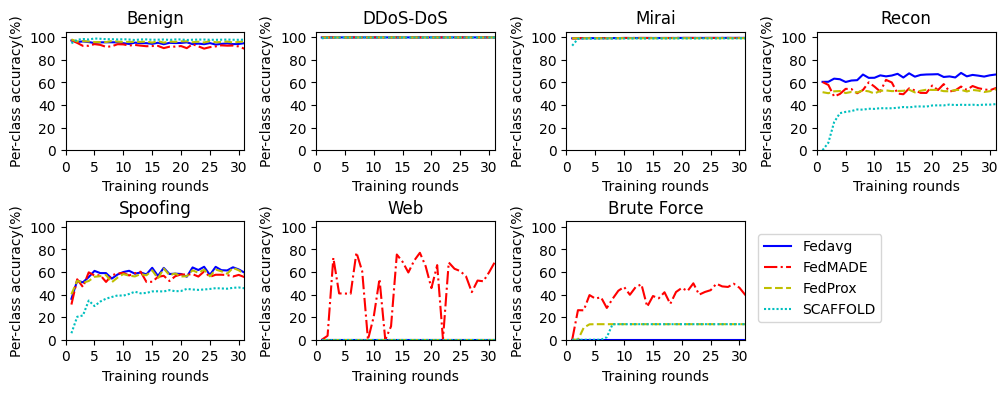}}
\caption{Curves of per-class accuracy across training rounds with $\gamma$ being 0.5.}
\label{fig:training_cnn_frac}
\vspace{-0.1in}
\end{figure*}

\textit{Training Process and Convergence Analysis}: Figures~\ref{fig:training_cnn} and~\ref{fig:training_cnn_frac} present the per-class accuracy during the training process. Despite differences in round training times, FedAvg, FedProx, and FedMADE demonstrate convergence within five training rounds. In contrast, SCAFFOLD exhibits a slower convergence rate compared to the other methods.
In Figure~\ref{fig:training_cnn_frac}, the per-class accuracy for web-based attacks rapidly decreases to zero in certain training rounds. This occurs because web-based attacks target only three out of 63 devices in the network. When these specific devices do not participate in a training round, the aggregated global model tends to forget the information associated with these attacks. However, by utilizing a validation dataset to evaluate the global model's performance in each communication round and preserving the best-performing model for deployment, the best model can ultimately achieve robust performance on minority attacks.

\bibliographystyle{splncs04}
\bibliography{main}

\begin{thebibliography}{10}
\providecommand{\url}[1]{\texttt{#1}}
\providecommand{\urlprefix}{URL }
\providecommand{\doi}[1]{https://doi.org/#1}

\bibitem{dpkt}
Dpkt library. \url{https://dpkt.readthedocs.io/en/latest/}, accessed: 2024-04-25

\bibitem{ML3}
Anthi, E., Williams, L., Słowińska, M., Theodorakopoulos, G., Burnap, P.: A supervised intrusion detection system for smart home iot devices. IEEE Internet of Things Journal  \textbf{6}(5),  9042--9053 (2019)

\bibitem{few_shot}
Ayesha~S., D., A.B., S., D., M.: Fs3: Few-shot and self-supervised framework for efficient intrusion detection in {Internet of Things} networks. In: Proceedings of the 39th Annual Computer Security Applications Conference (ACSAC). p. 138–149. Association for Computing Machinery (2023)

\bibitem{fed-dbn_2023}
Belarbi, O., Spyridopoulos, T., Anthi, E., Mavromatis, I., Carnelli, P., Khan, A.: Federated deep learning for intrusion detection in {IoT} networks. In: IEEE Global Communications Conference (GLOBECOM). pp. 237--242 (2023)

\bibitem{smote}
Chawla, N.V., Bowyer, K.W., Hall, L.O., Kegelmeyer, W.P.: Smote: synthetic minority over-sampling technique. Journal of artificial intelligence research  \textbf{16},  321--357 (2002)

\bibitem{dbscan}
Ester, M., Kriegel, H.P., Sander, J., Xu, X.: A density-based algorithm for discovering clusters in large spatial databases with noise. In: Proceedings of the Second International Conference on Knowledge Discovery and Data Mining (KDD). p. 226–231. AAAI Press (1996)

\bibitem{cluster_sample_2021}
Fraboni, Y., Vidal, R., Kameni, L., Lorenzi, M.: Clustered sampling: Low-variance and improved representativity for clients selection in federated learning. In: International Conference on Machine Learning (ICML). pp. 3407--3416 (2021)

\bibitem{gradient_inver}
Huang, Y., Gupta, S., Song, Z., Li, K., Arora, S.: Evaluating gradient inversion attacks and defenses in federated learning. In: Advances in Neural Information Processing Systems (NIPS). vol.~34, pp. 7232--7241 (2021)

\bibitem{scaffold}
Karimireddy, S.P., Kale, S., Mohri, M., Reddi, S., Stich, S., Suresh, A.T.: Scaffold: Stochastic controlled averaging for federated learning. In: International conference on machine learning (ICML). pp. 5132--5143 (2020)

\bibitem{iot-sn_2020}
Khoa, T.V., Saputra, Y.M., Hoang, D.T., Trung, N.L., Nguyen, D., Ha, N.V., Dutkiewicz, E.: Collaborative learning model for cyberattack detection systems in iot industry 4.0. In: IEEE Wireless Communications and Networking Conference (WCNC). pp.~1--6 (2020)

\bibitem{IDS_survey}
Khraisat, A., Gondal, I., Vamplew, P., Kamruzzaman, J.: Survey of intrusion detection systems: techniques, datasets and challenges. Cybersecurity  \textbf{2}(1),  1--22 (2019)

\bibitem{BoT-IoT}
Koroniotis, N., Moustafa, N., Sitnikova, E., Turnbull, B.: Towards the development of realistic botnet dataset in the {I}nternet of {T}hings for network forensic analytics: Bot-iot dataset. Future Generation Computer Systems  \textbf{100},  779--796 (2019)

\bibitem{MOON}
Li, Q., He, B., Song, D.: Model-contrastive federated learning. In: Proceedings of the IEEE/CVF conference on computer vision and pattern recognition (CVPR). pp. 10713--10722 (2021)

\bibitem{fedprox}
Li, T., Sahu, A.K., Zaheer, M., Sanjabi, M., Talwalkar, A., Smith, V.: Federated optimization in heterogeneous networks. Proceedings of Machine learning and systems (MLSys)  \textbf{2},  429--450 (2020)

\bibitem{AMCNN-LSTM_2021}
Liu, Y., Garg, S., Nie, J., Zhang, Y., Xiong, Z., Kang, J., Hossain, M.S.: Deep anomaly detection for time-series data in industrial {I}o{T}: A communication-efficient on-device federated learning approach. IEEE Internet of Things Journal  \textbf{8}(8) (2021)

\bibitem{Fedsld}
Luo, J., Wu, S.: Fedsld: Federated learning with shared label distribution for medical image classification. In: 2022 IEEE 19th International Symposium on Biomedical Imaging (ISBI). pp.~1--5 (2022)

\bibitem{feature_selection}
Maldonado, J., Riff, M.C., Neveu, B.: A review of recent approaches on wrapper feature selection for intrusion detection. Expert Systems with Applications  \textbf{198},  116822 (2022)

\bibitem{signature}
Masdari, M., Khezri, H.: A survey and taxonomy of the fuzzy signature-based intrusion detection systems. Applied Soft Computing  \textbf{92},  106301 (2020)

\bibitem{fedavg}
McMahan, B., Moore, E., Ramage, D., Hampson, S., y~Arcas, B.A.: Communication-efficient learning of deep networks from decentralized data. In: Artificial intelligence and statistics (AISTATS). pp. 1273--1282. PMLR (2017)

\bibitem{N-BaIoT}
Meidan, Y., Bohadana, M., Mathov, Y., Mirsky, Y., Shabtai, A., Breitenbacher, D., Elovici, Y.: {N-BaIoT—Network-based detection of IoT botnet attacks using deep autoencoders)}. IEEE Pervasive Computing  \textbf{17}(3),  12--22 (2018)

\bibitem{fl-lstm-gru_2022}
Mothukuri, V., Khare, P., Parizi, R.M., Pouriyeh, S., Dehghantanha, A., Srivastava, G.: Federated-learning-based anomaly detection for {I}o{T} security attacks. IEEE Internet of Things Journal  \textbf{9}(4),  2545--2554 (2022)

\bibitem{TON}
Moustafa, N.: A new distributed architecture for evaluating ai-based security systems at the edge: Network {TON\_IoT} datasets. Sustainable Cities and Society  \textbf{72},  102994 (2021)

\bibitem{UNSW-NB15}
Moustafa, N., Slay, J.: {UNSW-NB15}: a comprehensive data set for network intrusion detection systems. In: 2015 Military Communications and Information Systems Conference (MilCIS). pp.~1--6 (2015)

\bibitem{CICIoT2023}
Neto, E.C.P., Dadkhah, S., Ferreira, R., Zohourian, A., Lu, R., Ghorbani, A.A.: {CICIoT2023: A real-time dataset and benchmark for large-scale attacks in IoT environment}. Sensors  \textbf{23}(13) (2023)

\bibitem{diot_2019}
Nguyen, T.D., Marchal, S., Miettinen, M., Fereidooni, H., Asokan, N., Sadeghi, A.R.: D{\"i}ot: A federated self-learning anomaly detection system for {IoT}. In: 2019 IEEE 39th International conference on distributed computing systems (ICDCS). pp. 756--767. IEEE (2019)

\bibitem{poison}
Nguyen, T.D., Rieger, P., Miettinen, M., Sadeghi, A.R.: Poisoning attacks on federated learning-based iot intrusion detection system. In: Proc. Workshop Decentralized IoT Syst. Secur.(DISS). vol.~79 (2020)

\bibitem{fed-pca_2023}
Nguyen, T.A., He, J., Le, L.T., Bao, W., Tran, N.H.: Federated {PCA} on grassmann manifold for anomaly detection in {IoT} networks. In: IEEE Conference on Computer Communications (INFOCOM). pp. 1--10 (2023)

\bibitem{iot}
Ni{\v{z}}eti{\'c}, S., {\v{S}}oli{\'c}, P., Gonzalez-De, D.L.d.I., Patrono, L., et~al.: {Internet of Things (IoT)}: Opportunities, issues and challenges towards a smart and sustainable future. Journal of cleaner production  \textbf{274},  122877 (2020)

\bibitem{ML2}
Pajouh, H., Javidan, R., Khayami, R., Dehghantanha, A., Choo, K.: A two-layer dimension reduction and two-tier classification model for anomaly-based intrusion detection in {IoT} backbone networks. IEEE Transactions on Emerging Topics in Computing  \textbf{7}(02),  314--323 (2019)

\bibitem{fl-botnet_2022}
Popoola, S.I., Ande, R., Adebisi, B., Gui, G., Hammoudeh, M., Jogunola, O.: Federated deep learning for zero-day botnet attack detection in {IoT}-edge devices. IEEE Internet of Things Journal  \textbf{9}(5),  3930--3944 (2022)

\bibitem{malware_det_2022}
Rey, V., {Sánchez Sánchez}, P.M., {Huertas Celdrán}, A., Bovet, G.: Federated learning for malware detection in {IoT} devices. Computer Networks  \textbf{204},  108693 (2022)

\bibitem{FL_private}
Shi, S., Haque, M.S., Parida, A., Linguraru, M.G., Hou, Y.T., Anwar, S.M., Lou, W.: Harvesting private medical images in federated learning systems with crafted models. arXiv  (2024)

\bibitem{scale-mia}
Shi, S., Wang, N., Xiao, Y., Zhang, C., Shi, Y., Hou, Y.T., Lou, W.: Scale-mia: A scalable model inversion attack against secure federated learning via latent space reconstruction. arXiv  (2023)

\bibitem{targeted}
Sun, S., Sugrim, S., Stavrou, A., Wang, H.: Partner in crime: Boosting targeted poisoning attacks against federated learning. arXiv  (2024)

\bibitem{NSL-KDD}
Tavallaee, M., Bagheri, E., Lu, W., Ghorbani, A.A.: {A detailed analysis of the KDD CUP 99 data set}. In: 2009 IEEE Symposium on Computational Intelligence for Security and Defense Applications. pp.~1--6 (2009)

\bibitem{ratio_loss}
Wang, L., Xu, S., Wang, X., Zhu, Q.: Addressing class imbalance in federated learning. In: Proceedings of the AAAI Conference on Artificial Intelligence. pp. 10165--10173 (2021)

\bibitem{feco_2022}
Wang, N., Chen, Y., Hu, Y., Lou, W., Hou, Y.T.: {FeCo}: Boosting intrusion detection capability in {IoT} networks via contrastive learning. In: IEEE Conference on Computer Communications (INFOCOM). pp. 1409--1418 (2022)

\bibitem{node_selection}
Wu, H., Wang, P.: Node selection toward faster convergence for federated learning on non-iid data. IEEE Transactions on Network Science and Engineering  \textbf{9}(5),  3099--3111 (2022)

\bibitem{FedPAC}
Xu, J., Tong, X., Huang, S.L.: Personalized federated learning with feature alignment and classifier collaboration. In: The Eleventh International Conference on Learning Representations (ICLR) (2023)

\bibitem{fedmix}
Yoon, T., Shin, S., Hwang, S.J., Yang, E.: Fedmix: Approximation of mixup under mean augmented federated learning. arXiv  (2021)

\bibitem{mindfl}
Zhang, C., Wang, N., Shi, S., Du, C., Lou, W., Hou, Y.T.: Mindfl: Mitigating the impact of imbalanced and noisy-labeled data in federated learning with quality and fairness-aware client selection. In: IEEE Military Communications Conference (MILCOM). pp. 331--338 (2023)

\bibitem{FL_noniid}
Zhao, Y., Li, M., Lai, L., Suda, N., Civin, D., Chandra, V.: Federated learning with non-iid data. arXiv  (2018)

\bibitem{ML1}
Zolanvari, M., Teixeira, M.A., Gupta, L., Khan, K.M., Jain, R.: Machine learning-based network vulnerability analysis of industrial {Internet of Things}. IEEE Internet of Things Journal  \textbf{6}(4),  6822--6834 (2019)

\end{thebibliography}

\end{document}